\documentclass[graybox]{svmult}

\usepackage{makeidx}         
\usepackage{graphicx}        

\usepackage{url}
\usepackage[bottom]{footmisc}

\makeindex             

\begin{document}

\title*{Temporal networks of face-to-face human interactions}

\author{Alain Barrat
\and Ciro Cattuto
}

\institute{Alain Barrat \at 
Aix Marseille Universit\'e, CNRS UMR 7332, CPT, 13288 Marseille, France\\
Universit\'e du Sud Toulon-Var, CNRS UMR 7332, CPT, 83957 La Garde, France\\
Data Science Laboratory, ISI Foundation, Torino, Italy\\
\email{alain.barrat@cpt.univ-mrs.fr}
\and Ciro Cattuto \at  Data Science Laboratory, ISI Foundation, Torino, Italy\\
\email{ciro.cattuto@isi.it}}

\maketitle

\abstract{
The ever increasing adoption of mobile technologies and
ubiquitous services allows to sense human behavior at unprecedented
levels of details and scale. Wearable sensors are opening up a new
window on human mobility and proximity at the finest resolution of
face-to-face proximity. As a consequence, empirical data describing
social and behavioral networks are acquiring a longitudinal
dimension that brings forth new challenges for analysis and modeling.
Here we review recent work on the representation and analysis of
temporal networks of face-to-face human proximity, based on
large-scale datasets collected in the context of the SocioPatterns
collaboration. We show that the raw behavioral data can be studied
at various levels of coarse-graining, which turn out to be
complementary to one another, with each level exposing different features of the
underlying system. We briefly review a generative model of temporal
contact networks that reproduces some statistical observables.
Then, we shift our focus from surface statistical features to dynamical
processes on empirical temporal networks. We discuss how simple
dynamical processes can be used as probes to expose important features
of the interaction patterns, such as burstiness and causal
constraints. We show that simulating dynamical processes on empirical
temporal networks can unveil differences between datasets that
would otherwise look statistically similar. Moreover, we argue that, due to
the temporal heterogeneity of human dynamics, in order to investigate
the temporal properties of spreading processes it may be necessary to
abandon the notion of wall-clock time in favour of an intrinsic notion
of time for each individual node, defined in terms of its activity level.
We conclude highlighting several open research questions raised by the
nature of the data at hand.
}

\section{Introduction}
\label{sec:intro}
Although social relationships and behaviors are inherently dynamically
evolving, social interactions, represented by the paradigm of social
networks~\cite{Wasserman:1994} have long been studied as static entities,
mostly because empirical longitudinal data have been scarce~\cite{Padgett:1993,Lubbers:2010},
and often limited to relatively small groups of individuals.
As new technologies pervade our daily life,
digital traces of human activities are gathered at many
different temporal and spatial scales and for large populations,
and they promise to transform the way we measure, model and reason
on social aggregates~\cite{compsocsci,compsocsci2} and socio-technical systems~\cite{sociotechnical}
that combine social dynamics and computer-supported interaction mechanisms
(e.g., large scale on-line social networks like Twitter and Facebook).
It is important to remark that, from a methodological point of view,
data from technological and infrastructural proxies give access to
\textit{behavioral} networks defined in terms of the specific proxy at hand,
and not to \textit{bona fide} social networks.
Digital traces have been already used as proxies to study many specific aspects of human behavior,
such as geographic mobility~\cite{Chowell:2003,Montis:2007,brockmann,alain-vespi,Balcan:2009,Gonzalez:2008,Song:2010},
phone communications~\cite{Onnela:2007},
email exchange or instant messaging~\cite{Eckmann:2004,Kossinets:2006,Golder:2007,Leskovec:2008,Makse:2009,Amaral:2009},
and even human mobility and proximity in indoor environments~\cite{Sociopatterns,Cattuto:2010,alani,percol,Salathe:2010}.

Due to the often high temporal resolution of emerging data sets on human interactions,
the now customary representation of interactions in terms of static complex
networks, which has led to countless interesting analyses and insights~\cite{science,Dorogovtsev:2003,Newman:2003,Pastor:2004,Caldarelli:2007,Barrat:2008,Wasserman:1994,watts-short},
needs to be extended to take into account the dynamical properties of
the interaction patterns, bringing forth the field of ``temporal networks''~\cite{review_holme}.
This prompts fundamental and applied research on adapting and extending well-known
networks observables, metrics and characterization techniques to the
more complex case of a time-varying graph representations.

At the same time, the availability of high-resolution time-resolved data on human interactions
does not mean that any research question should be addressed by using
the full-scale and finest-resolution datasets, as they usually entail computational
challenges due to the sheer size of their digital representations.
Given a specific problem or research questions,
understanding what is the most appropriate scale for coarse-graining the raw
behavioral data, and what are the mathematical techniques and data representation that are best suited
to create such synopses is a key problem in its own merit~\cite{Clauset:2007,Caceres:2011,Krings:2012},
and has been already identified as such in the specific case of epidemic simulation
based on temporal social network data~\cite{Stehle:2011b,Blower2011}.

In the context outline above, we review here recent research efforts, developed
within the SocioPatterns collaboration~\cite{Sociopatterns}, based on
large-scale datasets that describe human face-to-face interactions in
various contexts, covering scientific conferences~\cite{Cattuto:2010,Isella:2011,Stehle:2011b,alani,percol},
hospital wards~\cite{Isella:2011b}, schools~\cite{Stehle:2011}, and museums~\cite{Isella:2011}.

\section{Phenomenology}
\label{sec:phenomenology}

\subsection{From raw proximity data to dynamical networks}
The data we describe here have been collected in deployments of
the SocioPatterns~\cite{Sociopatterns} social sensing infrastructure,
described in Ref.~\cite{Sociopatterns,Cattuto:2010}.
This measurement infrastructure is based on wearable wireless devices
that exchange low-power radio signals in a distributed fashion
and use radio packet exchange rates to monitor for location and proximity
of individuals. The proximity information is sent
to radio receivers installed in the environment, which timestamp and log contact data.
Participants are asked to wear such devices, embedded in unobtrusive wearable
badges, on their chests, so that badges can exchange radio packets
only when the individuals wearing them  face each other at close range
(about $1$ to $1.5$ m). The onboard software of the devices is tuned
so that the face-to-face proximity of two individuals wearing the badges
can be assessed with a probability in excess of $99\%$ over
an interval of $20$ seconds. A ``contact'' between two individuals is
then considered as established during a time period of $20$ seconds if
the devices worn by these individuals exchanged at least one
radio packet during that interval. The contact is then considered as ongoing
until a $20$ seconds interval occurs such that no packet exchange between the devices is recorded:
at that point the contact event is recorded together with its starting time and duration.
Notice that in contrast to other temporal network datasets in which interactions
are instantaneous events, here close-proximity and face-to-face contacts
do have a finite duration.

The data gathered by the social sensing infrastructure thus give access,
for each pair of participants, to the detailed list of their contacts,
with starting and ending times: these data can be 
represented as a time-varying social network of contact within the monitored community.
The temporal resolution of $20$ seconds in assessing proximity sets
the finest resolution for the temporal network representation we use,
which, in the following, will be assumed to be an ordered sequence
of graphs, each corresponding to a $20$-second interval.
Table~\ref{table:deployments} provides information and literature references
on the temporal networks that will be discussed in the following.

\begin{table}
	\centering
	\begin{tabular}{ |c|c|c|c|c|c|c|}
	\hline
	name & date & venue & event type & \# persons & duration & reference \\
	\hline
	\textbf{SG} & Apr-Jul 2009 &	Science Gallery, Dublin, IE & exhibition & $\sim 30,000$ & 3 months & \protect{\cite{infectious,Isella:2011,data:sg}} \\
	ESWC09 & Jun 2009	& ESWC 2009, Crete, GR & conference & $\sim 180$  & 4 days & \protect{\cite{alani,percol}} \\
	SFHH & Jun 2009	& SFHH, Nice, FR &  conference & $\sim 400$ & 2 days & \protect{\cite{Stehle:2011b}} \\
	\textbf{HT09} & Jul 2009 & ACM Hypertext 2009, Torino, IT &  conference & $\sim 120$ & 3 days & \protect{\cite{ht2009,data:ht09}} \\
	\textbf{PS} & Oct 2009 & primary school, Lyon, FR & school & $\sim 250$ & 2 days & \protect{\cite{Stehle:2011,data:school}} \\
	ESWC10 & Jun 2010 & ESWC 2010, Crete, GR & conference & $\sim 200$ & 4 days & \protect{\cite{iswc2010}} \\
	(OBG) & Nov 2009 & Bambino Ges\`u hospital, Roma, IT & hospital & $\sim 100$ & 10 days & \protect{\cite{Isella:2011b}} \\
	(PRAMA) & Apr 2010 & Practice Mapping, Gijon, ES & exhibition & $\sim 100$ & 10 days & \protect{\cite{prama}} \\
	(HFARM) & Jun-Jul 2010 & H-Farm, Treviso, IT & company & $\sim 200$ & 6 weeks & -- \\ 
	\hline
	\end{tabular}
	\caption{Partial list of the datasets on face-to-face
          proximity collected by the SocioPatterns collaboration
          during 2009 and 2010 and discussed in the present
          paper. Deployments that did not involve face-to-face
          detection or had less than $100$ participants are not
          reported. Deployments with names in bold face correspond to
          publicly available datasets (see references). Deployments
          with names in parentheses are listed for reference only.}
	\label{table:deployments}
\end{table}

\subsection{Microscopic view}
\label{sec:microscopic}
For each pair of individuals $i$ and $j$, the datasets contain a list
of $\ell$ successive time intervals $((t_{ij}^{(s,1)},
t_{ij}^{(e,1)}), (t_{ij}^{(s,2)}, t_{ij}^{(e,2)}), \cdots,
(t_{ij}^{(s,\ell)},t_{ij}^{(e,\ell)}))$ during which $i$ and $j$ were
detected to be in close-range face-to-face proximity, where $t_{ij}^{(s,a)}$ refers to the starting time (hence
the superscript $s$) and $t_{ij}^{(e,a)}$ to the ending time of the time interval number $a$.

Several quantities of interest can be defined to summarize the contact
patterns of each individual or pair of individuals, and to provide a
statistical characterization of contact patterns.
In particular, for each pair of individuals $i$ and $j$ (edge $i$-$j$), their list of contact time intervals
yields a list of contact durations $(\Delta t_{ij}^{(1)},\cdots,\Delta
t_{ij}^{(\ell)})$, with $\Delta t_{ij}^{(a)}= t_{ij}^{(e,a)}-t_{ij}^{(s,a)}$ for $a=1,\cdots,\ell$. 
Several notions of weight $w_{ij}$ for the edge $i$-$j$ can be defined
on the basis of this list of contact durations, yielding weighted contact networks
that describe different aspects of the empirical sequence of contacts:
\begin{itemize}
\item edge presence: $w^{p}_{ij}$ measures the contact occurrence (the superscript $p$ stands for ``presence''), with
$w^{p}_{ij}=1$ if at least one contact between $i$ and $j$ has been
established, and $0$ otherwise.

\item frequency of occurrence: the frequency $w^{n}_{ij}=l$ indicates
how many distinct contact events have been registered between $i$ and
$j$, disregarding the length of each contact (the superscript $n$ is for ``number'').

\item cumulative time in contact:  the cumulative duration of the contact
$w^{t}_{ij} = \sum_a \Delta t_{ij}^{(a)} $ gives the sum of the durations of all contacts
established between $i$ and $j$ (hence the superscript $t$).
\end{itemize}
At the level of each individual $i$, the above weights $w$ can be aggregated
over all individuals $j$ who had a contact with $i$, i.e., $s_{i}
=\sum_{j} w_{ij}$, yielding the following notions of node strength:
\begin{itemize}
\item $s_{i}^{p} =\sum_{j} w^{p}_{ij}$
gives the number of distinct individuals with whom $i$ has established
at least one contact; i.e., the degree $k_i$ of $i$ in the behavioral contact network.
\item $s_{i}^{n} =\sum_{j} w^{n}_{ij}$ indicates the
overall number of contacts in which $i$ has been involved.
\item the cumulative contact time $s_{i}^{t} =\sum_{j} w^{t}_{ij}$ corresponds
to the total sum of the duration of all contacts involving individual $i$
\footnote{Note that $s_{i}^{t}$ might be larger than the total
time during which $i$ has been in contact with any individual, as
$i$ could be in contact at the same time with more than one individual.}.
\end{itemize}
Of course all the quantities described above can be measured over the
whole duration of the deployment, or for a restricted duration (for
instance, one day, as is natural for conferences or the
museum deployment we describe below). The choice of the aggregation time
allows, for instance, to investigate the inter and intra-day
variability of interaction patterns among different individuals.

\begin{figure}[ht]
\centering
\includegraphics[width=0.85\columnwidth]{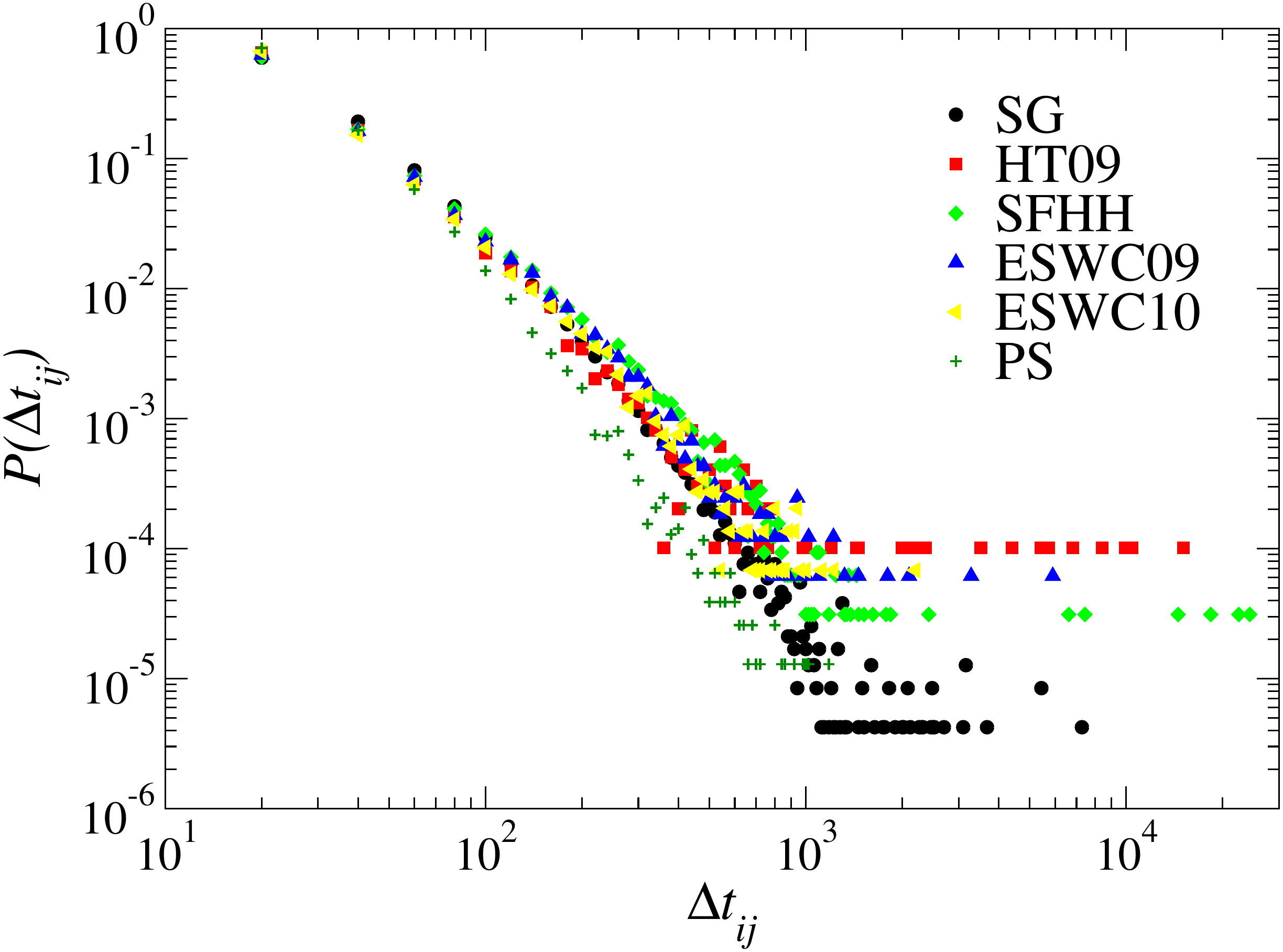}
\caption{Distributions of the face-to-face contact durations measured in different environments.}
\label{P-contact}
\end{figure}

\begin{figure}[htb]
\centering
\includegraphics[width=0.85\columnwidth]{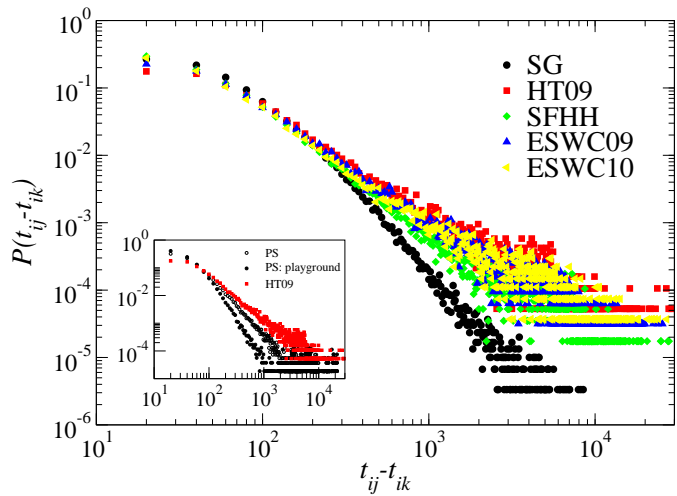}
\caption{Distributions of the time intervals between two successive
  contact events of a given individual, aggregated over the entire
  population.}
\label{intervals}
\end{figure}

A first way to uncover the complexity of the data is through a study
of the statistical distributions of the duration of the contact
events, and of the time intervals between contact events.
As shown in Figs.~\ref{P-contact} and \ref{intervals} and discussed in Ref.~\cite{Cattuto:2010},
broad distributions spanning several orders of magnitude are observed in both cases:
most contact durations and intervals between successive contacts are very short,
but very long durations are also observed, and no characteristic timescale emerges.
This bursty behavior is a well known feature of human dynamics and has been observed 
in a variety of systems  driven by human actions~\cite{humdyn-barabasi,humdyn-vazquez,Onnela:2007,Barabasi:2010}.
In the present case of close-range contacts, no simple functional form such as a power-law distribution
or a log-normal distribution seems to fit the observed data over the full range of time intervals.
However, it is important to highlight a few important aspects of the contact duration distributions
of Fig.~\ref{P-contact}.

The first observation is that the distributions approximately collapse
on one another regardless of the specific context they refer to. This
is remarkable as the datasets refer to very different social
environments: exhibitions (SG), where visitors stream along a
pre-defined path of a museum, academic conferences (HT09, ESWC09,
ESWC10), where the same tightly knit community shares a small number
of social spaces for several days and meets according to a predefined
schedule, large-scale conferences (SFHH) where many people do not know
each other and very different social spaces coexist, such as plenary
rooms and exhibition spaces. In the case of the primary school (PS), the distribution
is slightly narrower, possibly due to strong schedule constraints such as the duration
of breaks between lectures, or to the fact that young children tend to have less long 
face-to-face interactions than adults.
Regardless of all these social, spatial and demographic differences,
face-to-face contact behavior appears to obey the same bursty behavior
across all contexts. This is an important fact for modelers, as it
implies that processes relying on contact durations can be modeled by
plugging into the model the empirically observed distribution, assumed
to depend negligibly on the specifics of the contact situation being
modeled.

\begin{figure}[htb]
\centering
\includegraphics[width=0.9\columnwidth]{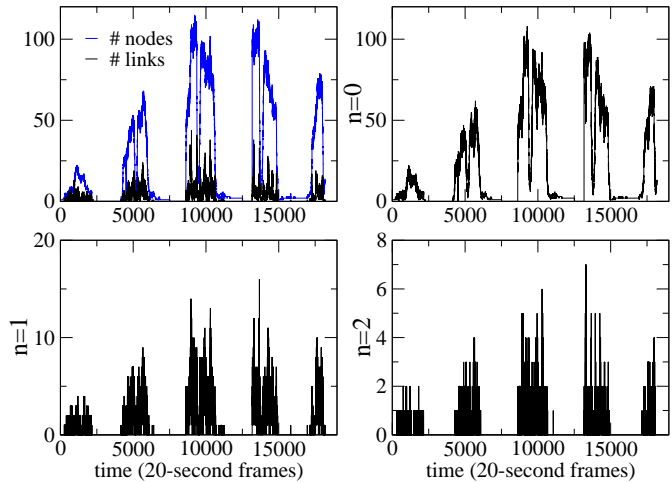}
\caption{Timelines of the number of: nodes and 
links in $20$-seconds instantaneous networks (top left), isolated nodes (top right), 
groups of 2 nodes (bottom left), groups of 3 nodes (bottom right) 
in the ESWC09 data set.
}
\label{timeline}
\end{figure}

A second observation deals with the origin of the contact duration heterogeneity of Fig.~\ref{P-contact}.
It may be argued that the simultaneous presence of multiple timescales of human contact
is responsible for the broad distribution we observe. However, as shown in Ref.~\cite{Cattuto:2010}
and related Supplementary Information, the contact durations restricted
to single individuals do exhibit the same broad distribution observed for the entire social aggregate.
This points to an intrinsic origin for the observed temporal heterogeneity, rooted in the way
single persons arbitrate their social contacts and the use of their time. 
The temporal heterogeneity of contact durations, at the individual and collective level,
undermines a number of simple representations for the contact network that implicitly
assume some degree of statistical homogeneity. For example, when dealing with
contact networks for epidemiological purpose, it is customary to summarize
the contact networks between classes of individuals by using contact matrices~\cite{Read:2012}
computed from averages of contact durations. Given the highly skewed character
of the actual distributions measured by using state-of-the-art techniques,
these representations need to be generalized in order to suitably capture
temporal and structural heterogeneities that may play  a crucial role
in determining the evolution of dynamical processes over contact networks.

\begin{figure}[htb]
\centering
\includegraphics[width=0.85\columnwidth]{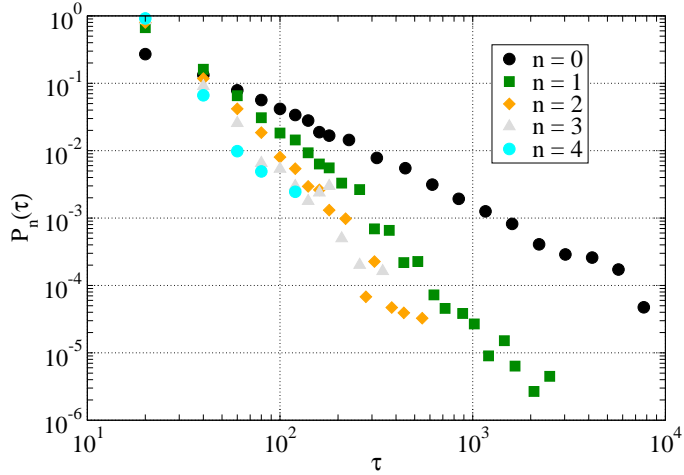}
\caption{Distributions of the lifetime (in seconds) of groups of size $n + 1$ (ESWC09).}
\label{groups}
\end{figure}

Perhaps the most striking feature of the observed distributions are their
robustness: the distributions of contact durations are extremely
similar for very different contexts, populations, activity timelines,
and deployment conditions. In particular, we observe the same
distribution in deployments corresponding to very different sampling
of the population under study (from $30\%$ for SFHH to almost $100\%$
for PS, HT09). This confirms the results of Ref.~\cite{Cattuto:2010} that
showed the robustness of the contact duration distribution under
further resampling of the data, aimed at simulating data loss or
limited population sampling. Moreover, although human activity and contact
patterns are highly non-stationary, as shown by an example in Fig.~\ref{timeline},
the contact duration distributions measured over different time
windows coincide~\cite{Cattuto:2010}, unveiling a statistical stationarity
in an otherwise non-stationary signal. This is consistent with similar analyses
on other temporal networks, such as proximity networks~\cite{Gautreau:2009}
and networks of cattle transfers between farms~\cite{Bajardi:2011}.

On the other hand, as displayed in Fig.~\ref{intervals},
the distribution of time intervals between successive contacts involving the same individual
typically do depend on the specific context at hand.
The distributons are very similar for different conferences (HT09, ESWC09, ESWC10, SFHH),
but narrower for the museum and the primary school cases.
Moreover, in the school case (inset in Fig.~\ref{intervals}),
spatial and behavioral sampling due to selecting only those contacts
that occur in the school playground significantly affects the distributions,
even though their qualitative features stay unchanged.
In general, this dependence on the context of the distribution of interval durations
between successive contacts means that, contrary to pair-wise interactions,
more complex temporal motifs that bear relevance to the causal structure
of the temporal network may depend on the specific environment.

Similar to the case of contact durations, broad distributions are also observed for the lifetimes
of simple structures in the contact network, such as groups of individuals
of size $n + 1$ ($n = 0$ corresponds to an isolated person, $n = 1$ to a pair of individuals, and so on),
as shown in Fig.~\ref{groups}. These broad distributions of group lifetimes 
become narrower for increasing $n$, i.e., larger groups are less stable than smaller ones.

\subsection{Aggregated network view}
\label{sec:aggregated}
The sequence of contact events between individuals during a given time
window defines an aggregated contact network at the population level,
which is a static summary of the temporal network.
In this network, each node is an individual, and a link between two nodes $i$ and $j$
denotes the fact that the corresponding individuals have been in
contact at least once during the time window under consideration.
Whereas the overall topological structure of the temporal network
can be encoded in a static graph, the temporal activity of individual edges
$i$-$j$ can be summarized by suitably defined weights for the edges,
such as the number of times $w^{n}_{ij}$ the link was established
or the cumulative duration $w^{t}_{ij}$ of the contact events between $i$ and $j$.

\begin{figure}[ht]
\centering
\includegraphics[width=0.85\columnwidth]{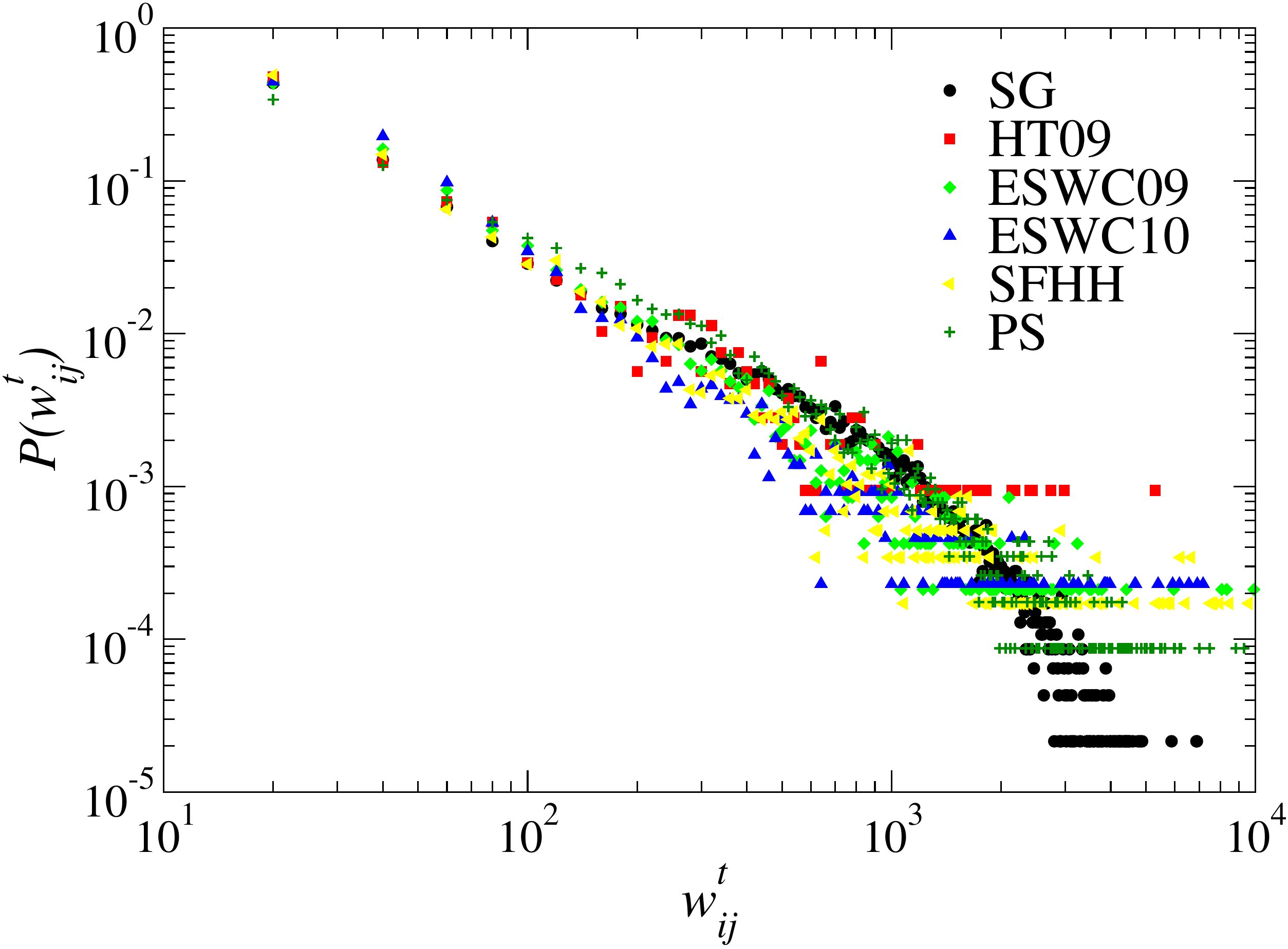}
\caption{Weight distributions for the daily aggregated networks. The weight $w^t_{ij}$ gives the cumulated amount
of time spent in face-to-face interaction during one day by individuals $i$ and $j$.}
\label{P-w}
\end{figure}  

The time window considered for aggregation can range from the finest
time resolution of $20$ seconds up to the entire duration of the data set.
In many contexts, it is natural to consider a specific temporal aggregation scale
(i.e., daily), but different aggregation levels typically provide complementary views
of the network dynamics at different scales.

Interestingly, and despite their static character, the structures of the aggregated
contact networks unveil important information about the contact
patterns of the population.
Let us first consider the statistical distributions that are typically
used to describe a network. The distributions of degree (number of
distinct individuals with whom a given individual has been in contact)
are typically narrow, with an exponential decay at large degrees
and characteristic average values that depend on the particular context~\cite{Isella:2011}.
On the other hand, the distributions of the cumulative contact durations
are broad: most pairs of individuals have been in face-to-face proximity
for a short total amount of time,  but a few cumulated contact durations are very long.
No characteristic interaction timescale can be naturally defined, except for obvious temporal cutoffs
due to the finite duration of the measurements.
As already observed in the case of the distributions of contact durations, 
Fig.~\ref{P-w} shows the similarity of the distributions
obtained in very different contexts: different populations,
in which individuals behave with very different goals in different spatial and social environments,
display a strikingly similar statistical behavior.

\begin{figure}[ht]
\centering
\includegraphics[width=0.4\columnwidth]{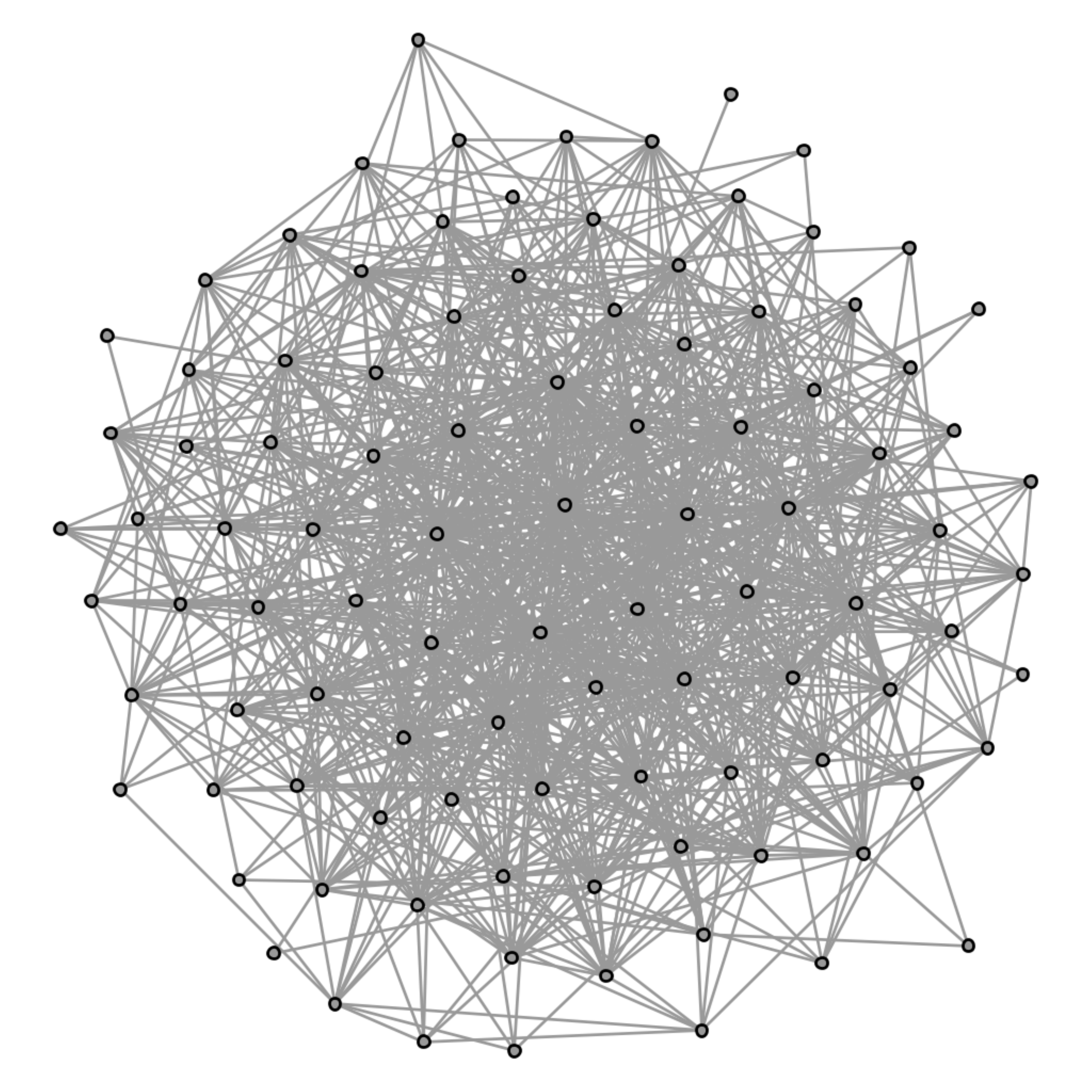}
\includegraphics[width=0.5\columnwidth]{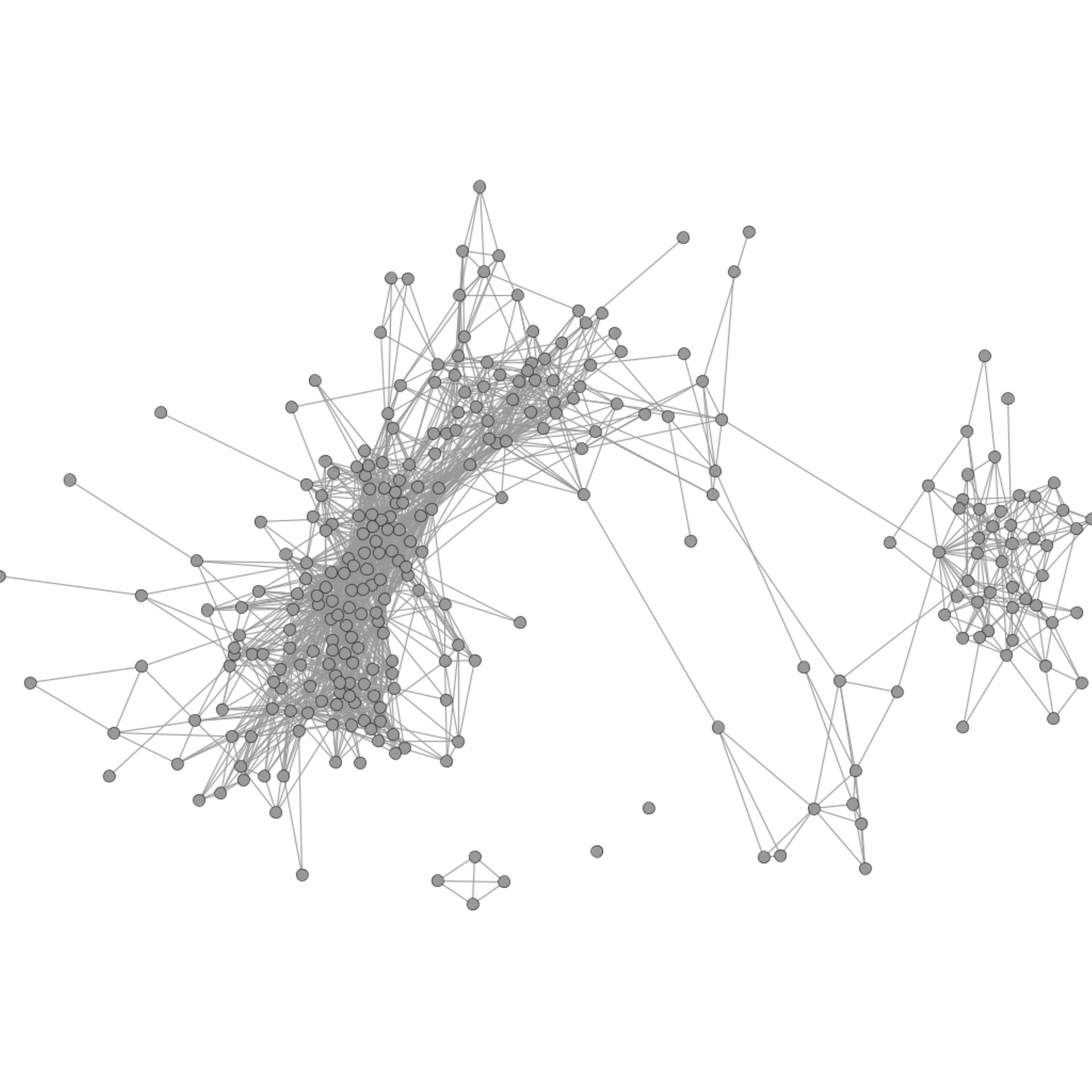}
\caption{Daily aggregated networks in the HT09 and SG deployments.
  Nodes represent individuals and edges are drawn between nodes if at
  least one contact event was detected during the aggregation
  interval. Left: aggregated network for one day of the HT09
  conference. Right: one representative day at the SG deployment. 
The network layouts were generated by using the force atlas graph
  layout implementation available in Gephi~\protect{\cite{gephi}}. }
\label{aggregated-networks}
\end{figure}

Despite their statistical similarities, the aggregated networks of
face-to-face proximity might have very different structures, as
revealed by a visual inspection of simple force-based network layouts.
For instance, Fig.~\ref{aggregated-networks} shows that
the aggregated network of interactions during a conference day
is much more ``compact'' than the ones describing the interactions
between museum visitors. In fact, as shown in Ref.~\cite{Isella:2011},
a typical daily aggregated network has a much smaller diameter in a conference context
than in the museum case. This difference is due to the different patterns of presence
of the attendees at the monitored venue, and also to the different social contexts:
in conferences participants are present during the entire conference duration
and are usually engaged in interacting with known individuals and in meeting
new persons. Conversely, the distribution of visit durations in the museum case
is close to a log-normal, with a geometric mean around $35$ minutes,
and the visitors' main goal is not to meet other visitors but rather to explore
the space following a partially pre-defined path. 
As a consequence, museum visitors are unlikely to interact directly
with other visitors entering the venue more than one hour after or before them,
thus preventing the aggregated network from having a short diameter:
there is limited interaction among visitors entering the museum at different times,
and the network diameter defines a path connecting visitors that
enter the venue at successive times, mirroring the longitudinal
dimension of the network. These findings show that aggregated network
topology and longitudinal/temporal behavioral properties are deeply interwoven.

\begin{figure}[ht]
\centering
\includegraphics[width=0.8\columnwidth]{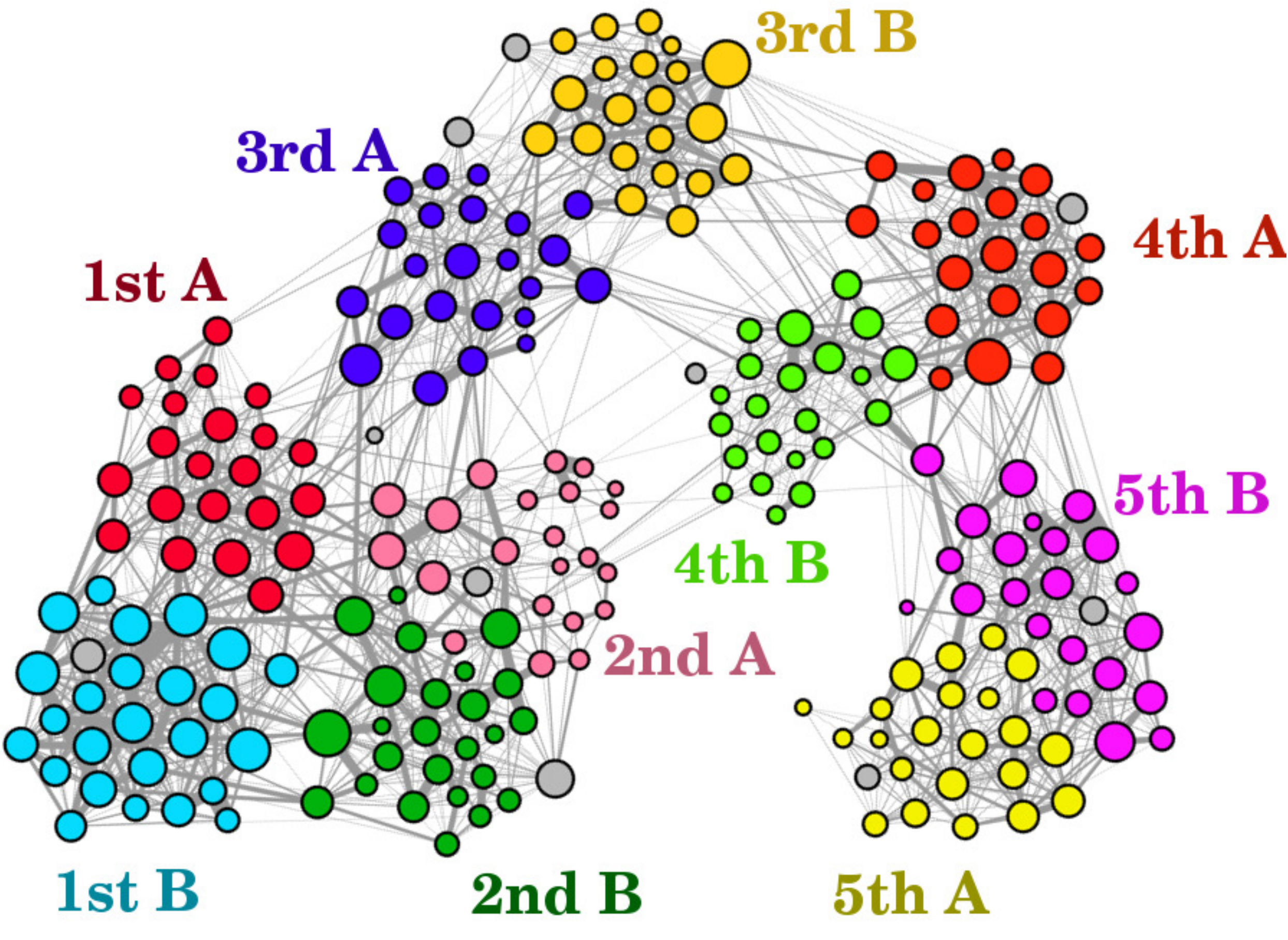}
\caption{School (PS) daily aggregated social network. Only links that
  correspond to cumulated face-to-face proximity in excess of $5$
  minutes are shown. The color of nodes indicates the grade and class
  of students. Grey nodes are teachers. The network layout was
  generated by using the force atlas graph layout implementation
  available in Gephi~\protect{\cite{gephi}}.}
  \label{aggregated-network_school}
\end{figure}

Finally, the aggregated network of contacts among school children
reported in Fig.~\ref{aggregated-network_school} represents an
intermediate case: children of each class form a cohesive structure
with many links, but links between different classes, and in
particular between children of different grades, are less frequent.
This structure results from several combined factors
that include 1) the spatial structure of the school,
the grouping of students into classes, 2) the fixed association
of school classes with given room for school activity,
3) the particular schedule of the school, according to which students
do not go to the schoolyard or canteen at the same time and their movement
as a group follows predefined spatio-temporal trajectories~\cite{Stehle:2011},
4) age-related homophily effects~\cite{McPherson:2001}, which also play a role
when children are free to move in the same space, such as in the playground.
Overall, as shown in Fig.~\ref{aggregated-network_school}
the cumulative contact network displays a visible community structure
that can be recalled using standard community-detection techniques.
However, it is important to remark that the cumulated network
projects out a lot of information on temporal communities,
i.e., on nodes that share similar spatio-temporal trajectories
and similar contact histories. For example, if a group $X$ of nodes mixes strongly
with a different group $Y$ during a time interval $[t_1,t_1+T]$,
and the second group of nodes $Y$ has some connections with a third group
$Z$ during $[t_2,t_2+T]$, with $t_2 > t_1+T$,
the cumulative network representation will lose information on the group
identities of $X$ and $Y$ and only show a single group $X \cup Y$
with connections to $Z$. The problem of defining and identifying
temporal communities in a time-varying networks~\cite{Tantipathananandh:2007,Seifi:2013} is a central one
when trying to mine out an activity schedule (such as the school schedule)
from electronic records of human interactions, and requires to suitably
define null models for temporal networks that incorporate the above described
heterogeneities.

These few examples show how similar statistical properties in terms of
heterogeneity of contact event durations and overall face-to-face
presence can in fact hide very distinct structures of aggregated
networks, that are shaped by the dynamical unfolding of the contacts:
the study of static aggregated networks sheds in this respect some light on
the system's dynamics.

\begin{figure}[ht]
\centering
\includegraphics[width=0.85\columnwidth]{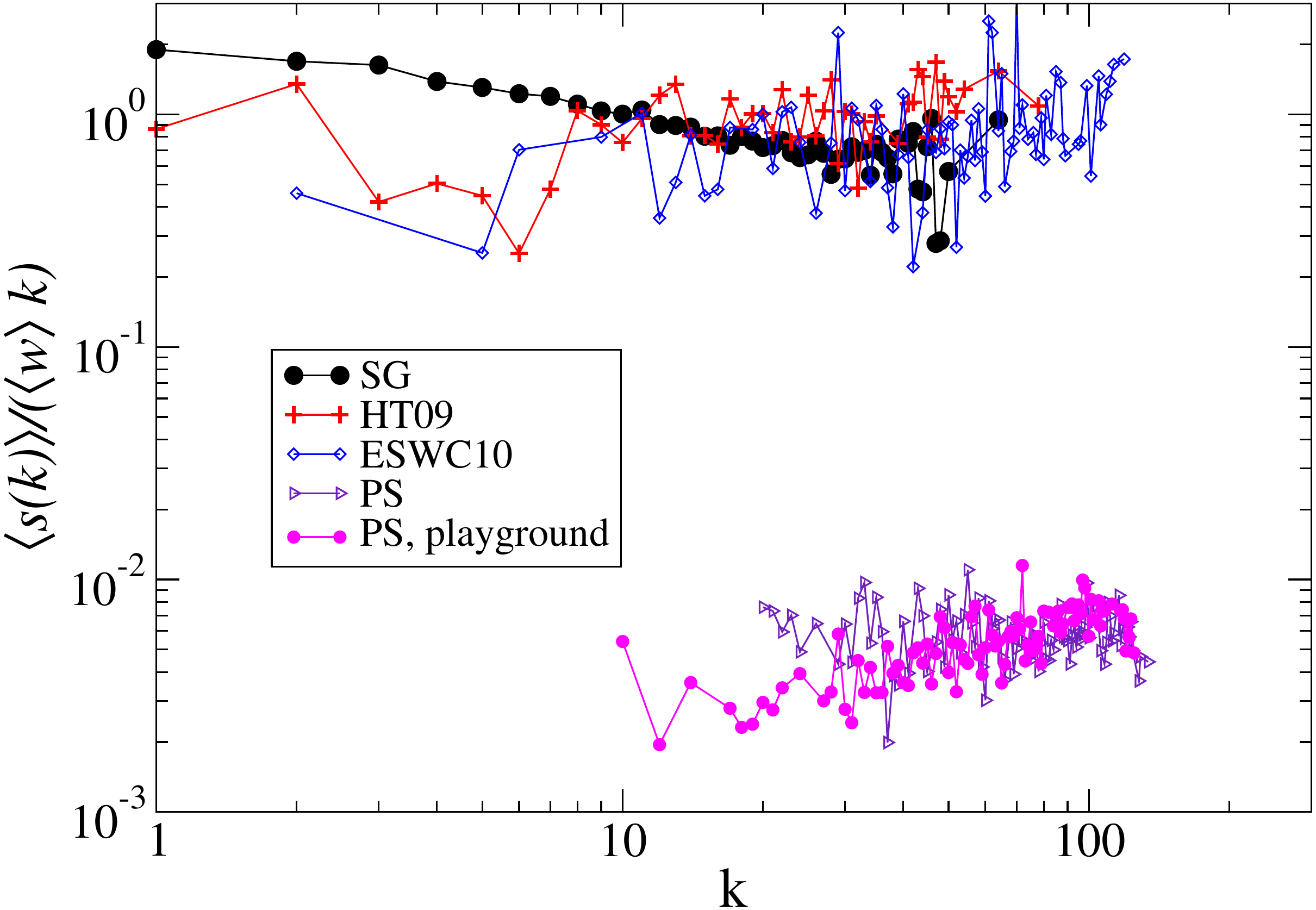}
\caption{Correlation between node's strength and degree, as measured
  by the average strength $\langle s(k) \rangle$ of nodes of degree
  $k$.  The figure shows $\langle s(k) \rangle / (\langle w \rangle
  k)$, in several contexts.  Distinct increasing and decreasing trends
  are observed, depending on the context.  }
\label{s-over-k}
\end{figure} 

As in other cases of weighted networks~\cite{alain-vespi},
more insight can be gleaned by studying the \textit{correlations} between the
weights, which are here the trace of the contact dynamics, and the topology.
Let us consider the strength $s$ of each node,
defined as the sum of the weights of all links inciding on it~\cite{alain-vespi}.
In our case, this corresponds, for each individual,
to the cumulated time of interaction with other individuals.
In social interaction contexts, it can be considered as at least as important as
the number of distinct individuals contacted (the degree in the
aggregated network), as it is a measure of the resources (time)
an individual committed to social interactions.
Correlations between the strength and the degree
are of course expected: even for completely random durations of the
contact events, a linear dependency of the average strength $\langle
s(k)\rangle$ of nodes of degree $k$ is obtained,
with $\langle s(k)\rangle \sim \langle w\rangle k$,
where $\langle w \rangle$ is the average link weight.
A deviation of $\langle s(k)\rangle/k$ from a horizontal line
thus denotes the existence of non-trivial correlations:
for instance, a decreasing $\langle s(k)\rangle/k$,
as observed in large-scale phone call networks~\cite{Onnela:2007},
indicates that individuals who call more distinct individuals spend on
average less time in each call than individuals who have less links.

In the face-to-face behavioral networks we review here,
two distinct behaviors have been observed, depending on the context,
as shown in Fig.~\ref{s-over-k}.
In the museum data set (SG), $\langle s(k) \rangle/(\langle w
\rangle k)$ has a clearly decreasing trend (that can be fitted by a
power law with a negative exponent). On the other hand, for the aggregated
networks describing the contacts in conferences (HT09 and ESWC10),  $\langle s(k) \rangle/(\langle w
\rangle k)$ displays consistently an increasing trend.
In school settings, a rather flat $\langle s(k) \rangle/(\langle w \rangle k)$
is observed when the contacts occurring in the whole school (PS) are taken
into account. This behavior has also been observed independently~\cite{Smieszek}, in another
dataset describing the proximity patterns of highschool students~\cite{Salathe:2010}.
However, if only the contacts occurring in contexts where the children's
movements and contacts are not constrained (PS, playground) are considered,
an increasing trend is found.

The contrasting results obtained in different contexts show that
processes such as information diffusion~\cite{vespi-classic,anderson-may},
frequently occurring in social contexts, will unfold in different ways.
The number of distinct persons encountered does not contain enough information
to estimate the spreading potential of an individual: a super-linear
dependence of $\langle s(k) \rangle$ with $k$ hints at the importance
of ``super-spreader nodes'' with large degree~\cite{vespi-classic,anderson-may}
while a sub-linear behavior indicates that the decrease in the weights of individual contacts
mitigates the expected super-spreading behavior of large degree nodes.

\section{Modeling face-to-face dynamical contact networks}
\label{sec:modeling}
The phenomenology outlined in the previous sections calls for the
development of new modeling frameworks for dynamically evolving
networks, as most modeling efforts have been until recently devoted to
the case of static networks
\cite{Dorogovtsev:2003,Newman:2003,Pastor:2004,Caldarelli:2007,Barrat:2008}.
Among recent models of dynamical networks
\cite{Gross:2008,Scherrer:2008,Hill:2009,Gautreau:2009,Stehle:2010,Stehle:2011c},
we review here a model of interacting agents developed in
Refs.~\cite{Stehle:2010,Stehle:2011c} in order to describe how individuals
interact at short times scales.

The model considers $N$ agents who can either
be isolated or form groups (cliques). Each agent $i$ is characterized by two
variables: (i) his/her  coordination number $n_i = 0, 1, 2 . . . , N-1$ 
indicating the number of agents interacting with him/her, and
(ii) the time $t_i$ at which $n_i$ was last changed.
At each (discrete) time step, an agent $i$ is chosen randomly. With a
probability $p_n(t,t_i)$ that may depend on the agent's state, on the
present time $t$ and on the last time $t_i$ at which $i$'s state
evolved, $i$ updates his/her state, under the following rules:
\begin{itemize}
\item[(i)]~If $i$ is isolated ($n_i=0$), another isolated agent $j$ is
  chosen with probability proportional to $p_0(t, t_j)$, and $i$ and
  $j$ form a pair. The coordination number of both agents are then
  updated ($n_i \rightarrow 1$ and $n_j \rightarrow 1$).

\item[(ii)]~If the agent $i$ is already in a group,
  ($n_i=n>0$), with probability $\lambda$ the agent $i$ leaves the
  group; in this case, the
  coordination numbers are updated as $n_i \rightarrow 0$, 
and $n_k \rightarrow n-1$ for all the other agents $k$ of the original
group. With probability $(1-\lambda)$ on the other hand $i$ introduces an isolated
  agent $j$ in the group, chosen
  with probability proportional to $p_0(t,t_j)$. The coordination
  numbers of all the interacting agents are then changed according to the
  rules  $n_j \rightarrow n+1$ and $n_k \rightarrow n+1$ (for all $k$ in the group).
\end{itemize}

These rules define a dynamic network of contacts between the agents,
whose properties depend on the probabilities $p_n$, which control the
tendency of the agents to change their state, and on the parameter
$\lambda$, which determines the tendency either to leave groups or on
the contrary to make them grow.

Constant probabilities $p_n$ correspond to Poissonian events of
creation and deletion of links between individuals, and hence to
narrow distributions of contact times. On the other hand, 
the introduction of memory effects in the definition of the $p_n$ 
is able to generate dynamical contact networks with properties similar
to the ones of empirical data sets \cite{Stehle:2010,Stehle:2011c}.
In particular, a reinforcement principle can be implemented by considering that the
probabilities $p_n(t,t')$ that an agent with coordination number $n$
changes his/her state decrease with the time elapsed since his/her last
change of state. To this aim, we can impose $p_n(t,t')=p_n(t-t')$, with  $p_n$
decreasing functions of their arguments. This is equivalent to the assumption 
that the longer an agent is interacting in a group, the smaller is the
probability that s/he will leave the group, and that the longer an agent is
isolated, the smaller is the probability that s/he will form a new
group. 

\begin{figure}
\centering
\includegraphics[width=0.8\textwidth]{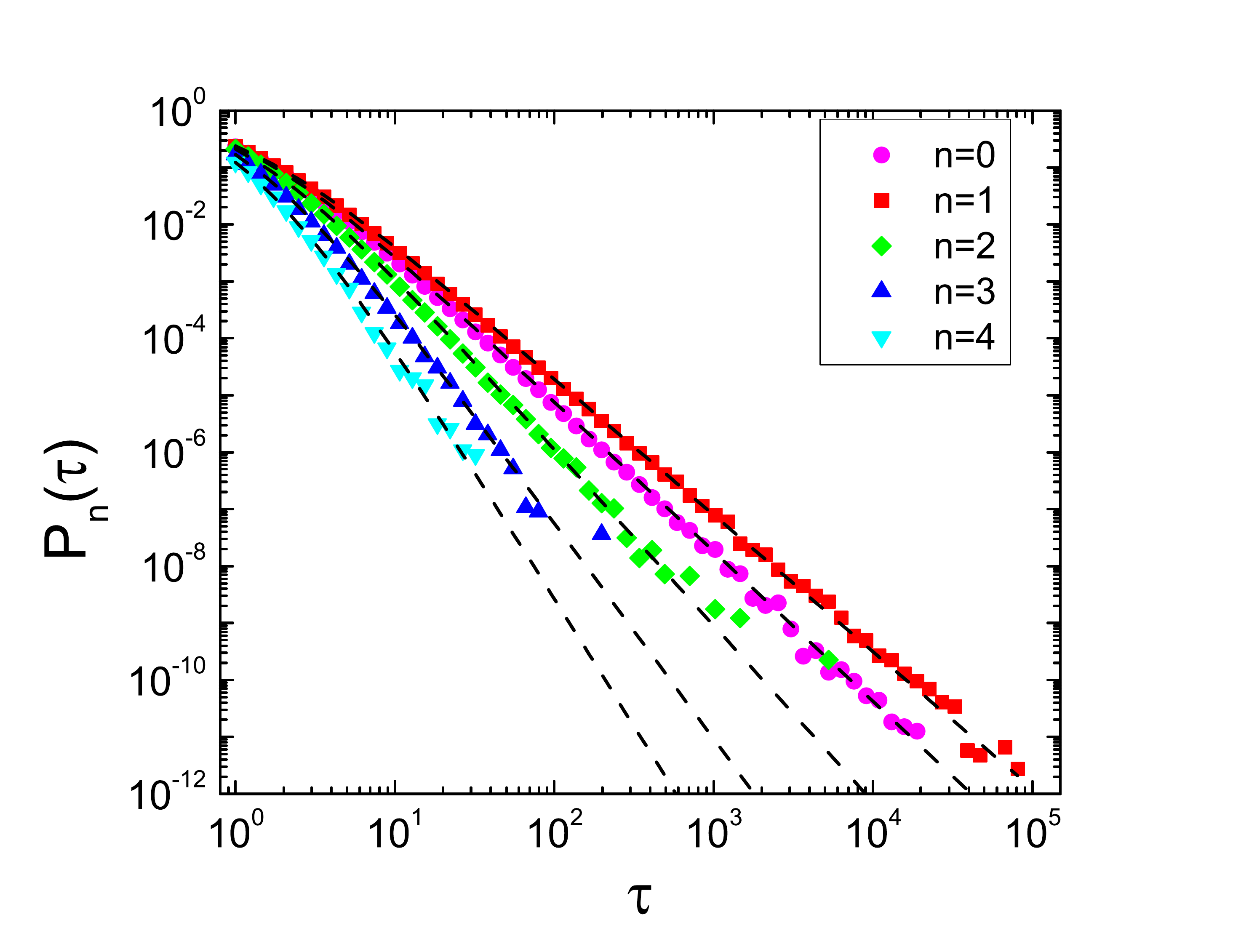}
\caption{Distribution $P_n(\tau) $ of durations of
  groups of size $n+1$, for $N=1000$ agents with $b_0=b_1=0.7$,
  $\lambda=0.8$, and a number of time steps
  $T_{max}=N\times 10^5$. The
  dashed lines correspond to the analytical predictions
  Eqs. (\ref{Pn2}). From \protect\cite{Stehle:2011c}.}
\label{Groups_stationary}
\end{figure}

The evolution equations of the number of agents in
each state can be solved self-consistently at the mean-field level \cite{Stehle:2010,Stehle:2011c}
in various cases. One of the simplest is
given by $p_n$ functions scaling as $1/(t-t')$, $
p_n(t,t')=\frac{b_n}{1+(t-t')/N}$, with 
moreover $b_n=b_1$ for every $n\geq1$
in order to reduce the number of parameters 
(the model's parameters are then $b_0$, $b_1$, and $\lambda$).
The probability distributions of the time spent in each state
can then be shown to be given by power-law distributions
\begin{eqnarray} \nonumber
P_0(\tau)&\propto &\left(1+\tau \right)^{-1-b_0(3\lambda-1)/(2\lambda-1)} \\
P_n(\tau) &\propto &\left(1+\tau \right)^{-(n+1)b_1-1} \ \mbox{for} \ n\geq 1. \label{Pn2}
\end{eqnarray}
As shown in Fig. \ref{Groups_stationary}, these predictions are
confirmed by numerical simulations~\footnote{The system is in a stationary
state for $b_1>0.5$, $b_0>(2\lambda-1)/(3\lambda-1)$ and
$\lambda>0.5$, while the self-consistent solution breaks down outside
of this parameter region, and non-stationary behavior with the
possible formation of large (system-size) groups can be observed
\cite{Stehle:2010,Stehle:2011c}.}. The distributions of time
intervals between successive contacts of an individual are as well
power-law distributed, and the aggregated contact networks display
features similar to the empirically observed ones.

The model can be easily extended to include an
intrinsic heterogeneity between agents, possibly reflecting a
difference in their ``sociability'', or to model populations with a
varying number of agents. For instance, it is possible to consider a
museum-like situation in which agents enter the system, remain for a
certain duration, and then leave without the possibility to re-enter
it. Power-law distributions of contact durations are
still observed, and the shape of the aggregated contact network
closely resembles the ones observed in the museum setting.
In addition, it would be possible to consider agents belonging to
different groups with different mixing properties, in order to mimic
as well for instance the contact dynamics in a school. Overall, this
model's versatility makes it an interesting tool for generating
artificial dynamical contact networks.

\section{Dynamical processes on dynamical networks}
\label{sec:dynamical}
Many networks are the support of dynamical processes of various nature,
from random walks to synchronization or spreading phenomena~\cite{Barrat:2008}.
Most related studies have however considered, as a first approach,
dynamical phenomena unfolding on static networks.
It has been shown how different topological characteristics
impact the unfolding of phenomena such as epidemic spreading,
with important consequences such as the suppression of the epidemic threshold
in very heterogeneous networks~\cite{vespi-classic}.

To date, few research efforts have dealt with the fact that the networks supporting
these phenomena might have a dynamics of their own.
Through the study of toy models of co-evolution, in which the network dynamics itself
is defined as a reaction to the process unfolding on top of it, it was shown that the interplay
of these two dynamics can lead to interesting and sometimes counter-intuitive effects~\cite{Gross:2008}.
In this context, the study of dynamical processes on dynamical (temporal) networks
can have a two-fold purpose.
On the one hand, it can guide the design of more realistic models,
applied for instance to epidemic spreading. Dynamical processes on temporal networks
can be studied to understand the relative roles of the different time scales at play,
and what level of  information is actually needed for the description of these processes~\cite{Stehle:2011b}.
In addition to that, the use of very simple dynamical processes, such as random walks or deterministic
spreading, can be considered among the techniques developed for the study and characterization
of  dynamical networks, such as the ones put forward in Refs.~\cite{Isella:2011,Bajardi:2011,Nicosia:2011,kovanen,review_holme}.
In particular, dynamical processes can serve to probe the role of causality constraints
in temporal networks, comparing the outcome of a given process
(i) on the dynamical network describing the real temporal sequence of events, 
and (ii) on aggregated networks in which the information about the precise order of events is discarded.

In this section we focus on a simple snowball SI model
of epidemic spreading or information diffusion~\cite{anderson-may}.
Individuals can be either in the susceptible (S) state,
indicating that they have not been reached by the ``infection'' (or information) yet,
or they can be in the infectious (I) state, meaning that they have been infected
by the disease (or that they have received the information)
and can further propagate it to other individuals.
In the simplest, deterministic version of  such a process,
every contact between a susceptible individual and an infectious one results
in a transmission event,  which instantaneously turns the susceptible individual
into an infected one according to $S+I \to 2I$.
In this simple model, infected individuals do not recover,
i.e., once they transition to the infectious state they indefinitely remain in that state.
The process is initiated by a single infected individual (``seed''), typically chosen at random.
Despite its simplicity, such a schematic model provides interesting insight
on causality constraints in dynamical networks,
and on how different temporal contact patterns can lead to different outcomes.
This is partially due to the fact that the SI process yields the fastest possible
information propagation from the seed node to the rest of the network,
thus bounding other more complex and realistic invasion processes.

\subsection{SI model as a probe of temporal network structure}
\label{sec:si}
\begin{figure}[ht]  
\includegraphics[width=0.5\columnwidth]{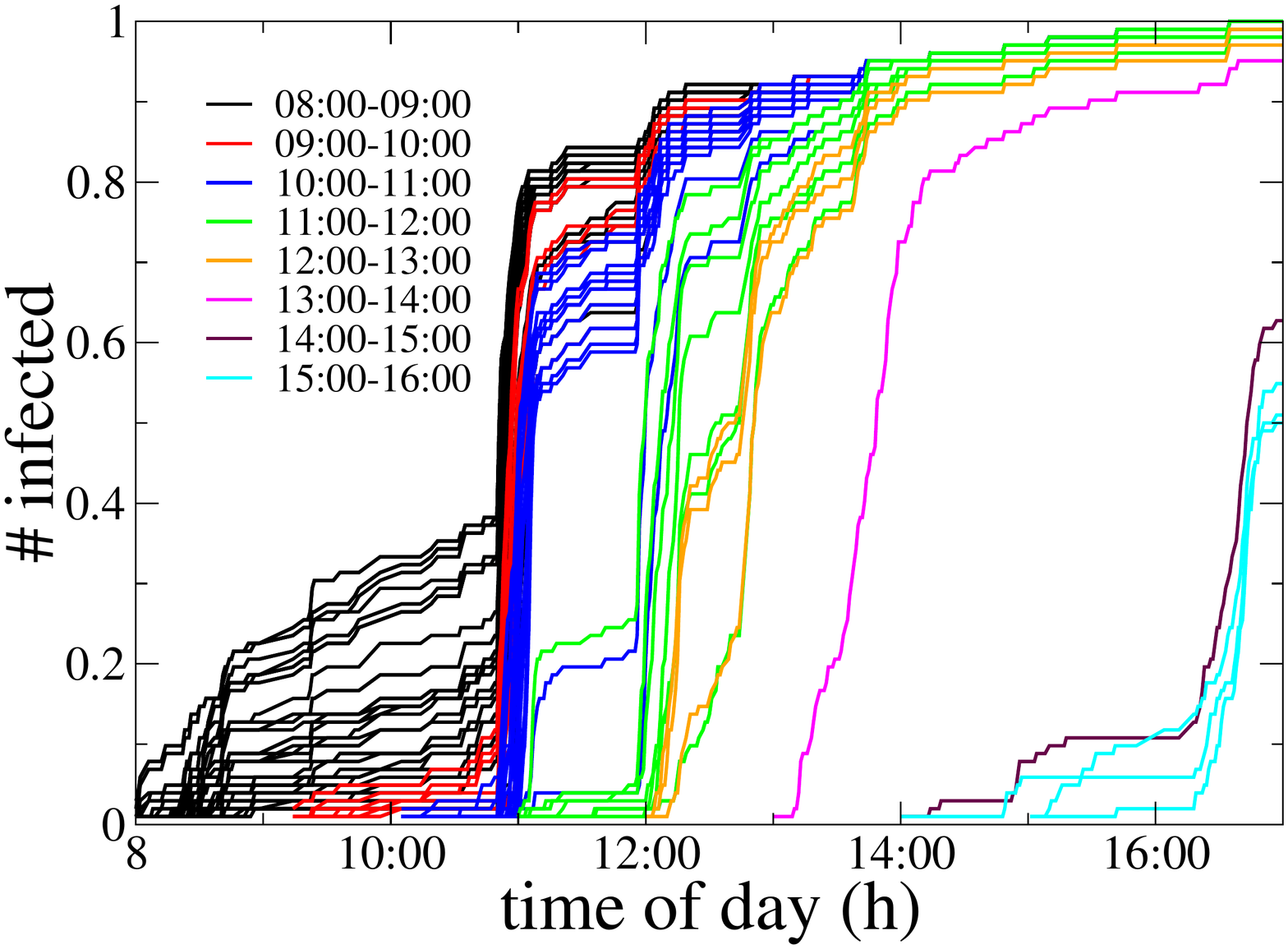}
\includegraphics[width=0.5\columnwidth]{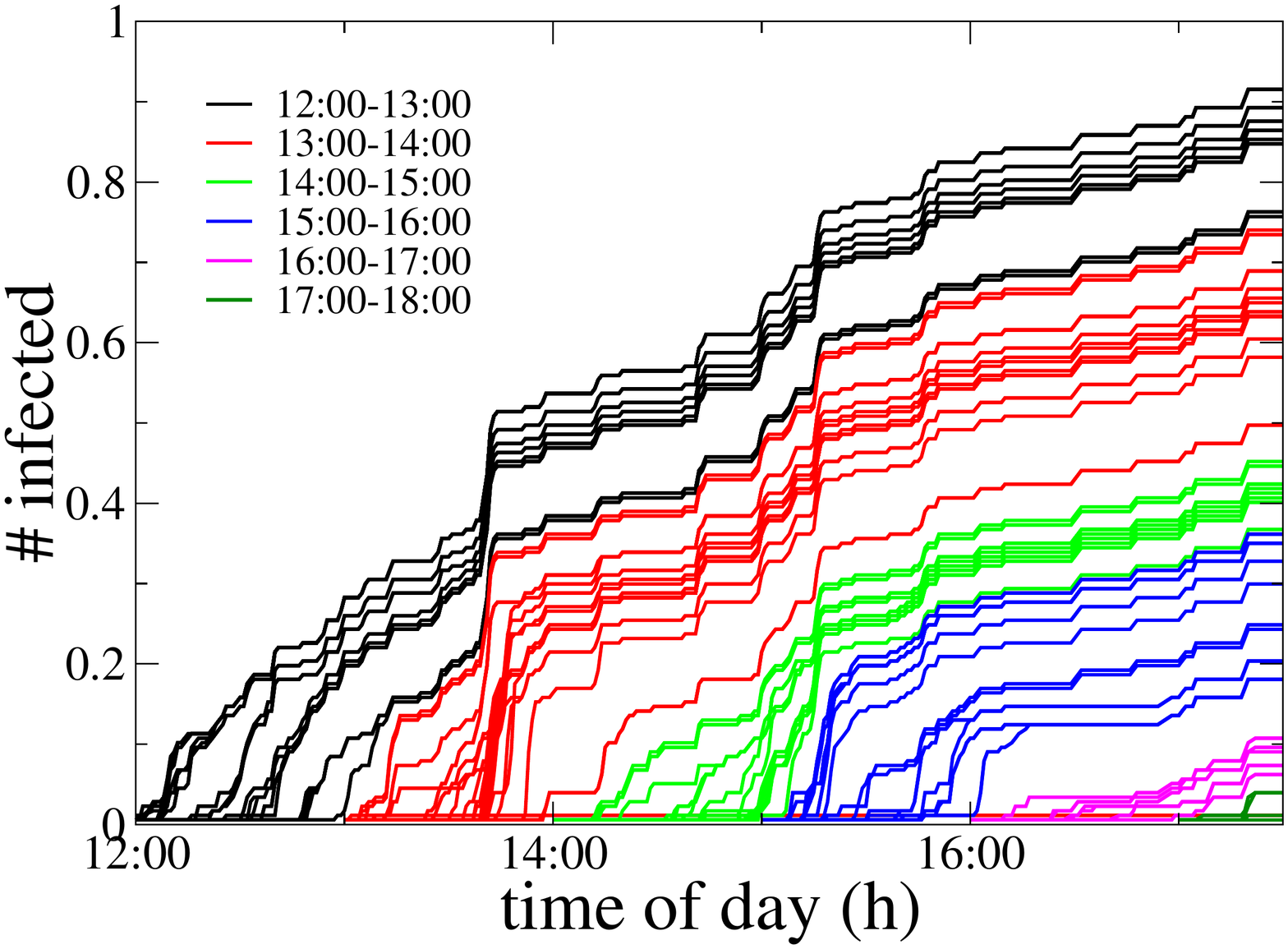}
\includegraphics[width=0.5\columnwidth]{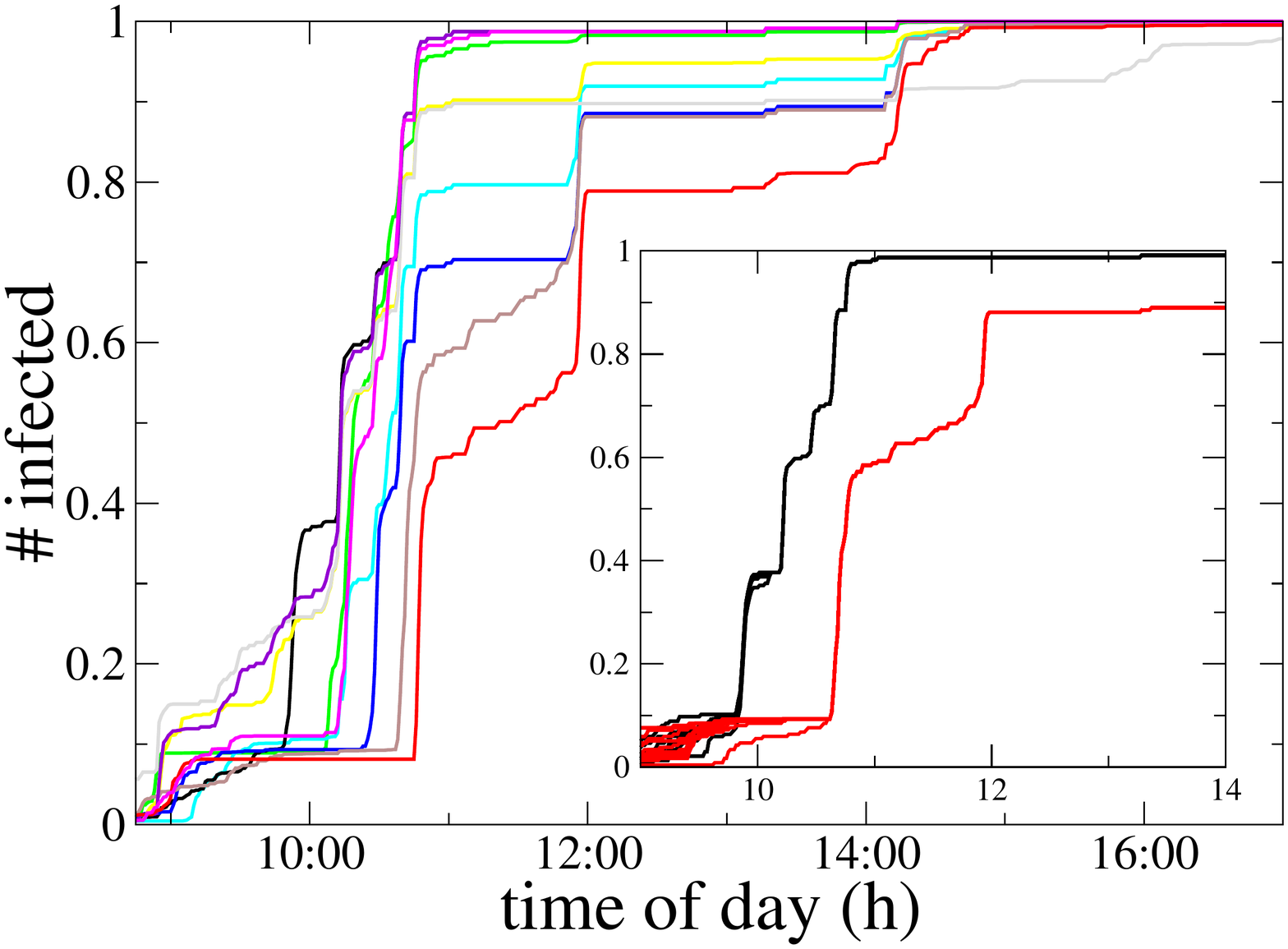}
\caption{Incidence curves showing the number of infected individuals as a function of time
for a susceptible-infectious (SI) process simulated over $1$ day of HT09 (top-left panel),
SG (top-right panel) and PS (bottom-left panel) temporal network data.
In the HT09 and SG cases (top plots) each curve corresponds to a different seed node
and is color-coded according to the starting time of the spreading process.
In the PS case (bottom) school, each curve is an average over the dynamics simulated
for different seed nodes belonging to a given class. The inset shows individual curves
for different seeds chosen in two distinct classes.}
\label{epidemics-within-day}
\end{figure}

Let us first consider the simplest measure of the unfolding of an SI
process in a population, as given by the temporal evolution
of the number of infected individuals (i.e., the incidence curve). 
Figure~\ref{epidemics-within-day} shows that the incidence curves
in different environments look qualitatively very different.

In the case of a typical day at a conference (HT09, top-left panel)
few transmission events occur until the conference participants gather
for the coffee break at $11$am, even if the seed was present early.
A strong increase in the number of infected individuals is then observed,
and a second strong increase occurs during the lunch break ($12$pm).
As a result of the concentration in time of transmission events,
spreading processes initiated by different seeds all achieve very similar
(and high) incidence levels after a few hours,
regardless of the initial seed and of its arrival time.
Even epidemics started by latecomers can reach about $80\%$ of the community.

A very different picture is observed in the museum case (top-right panel),
where the arrival time of the seed individual has a strong impact on the epidemic size:
at any point in time, the number of infected individuals is strongly correlated
with the arrival time of the seed. This is due to the fact that visitors stream through
the venue, and those who left before the arrival time of the seed 
cannot be reached by the infection.
Furthermore, in many cases the daily aggregated network displays
multiple disconnected components, so that the spreading process stops soon after
the seed leaves and only reaches a very small portion of the total population~\cite{Isella:2011}.
Even on days with many visitors and a globally connected daily aggregated network,
the incidence curves do not present sharp gradients as in the conference case,
and later epidemics are unable to infect a large fraction of daily visitors.

In the school case (bottom-left panel) almost all students arrive at the beginning
of the day, hence no effects due to heterogeneous arrival times can be observed.
Similarly to the conference case,
the simulated dynamics displays jumps in the number of infected individuals
at specific times of the day, regardless of the seed node,
and by the end of the day almost all individuals get infected.
Heterogeneous invasion dynamics can be observed depending
on the class of the seed node, and the differences can be related
to the scheduled activities and movements of school children.
Conversely, different choices for the seed node within a single class
(bottom-left panel, inset) yield very similar incidence curves.
This can be understood as a result of different within-class and cross-class
contact patterns: contacts within individual classes are rather homogeneous
compared to cross-class contacts, so that the SI process quickly reaches
most nodes of the seed's class. The invasion of other classes
is controlled by slower temporal structures
of the contact network, which are determined by the school schedule
and determine the sensitivity on the initial class.

To characterize in a more quantitative fashion
the importance of the temporal structure of the network on the spreading dynamics,
we can consider the number of individuals who are reachable
from the seed node through paths in the (daily) cumulative contact network,
and compute what fraction of them are actually reached by an SI process
that takes place over the temporal contact network.
The value of this ratio displays markedly different distributions
in the museum and in the conference case~\cite{Isella:2011}:
at a conference almost all the individuals who are reachable along
the cumulative contact network always get infected by the end of the day,
whereas in the museum case this ratio is often much smaller than $1$.
Therefore, studying the dynamics of a simple SI process can uncover
differences in the temporal structure of human proximity networks
that cannot be detected by using simpler statistical indicators
(e.g., the probability distributions of contact durations).
This calls for more work aimed at using generic dynamical processes
over temporal networks to define a new class of time-aware
network observables that can expose important differences and similarities.

\subsection{Causality-respecting paths}
\label{sec:paths}
The differences in spreading patterns outlined above are due
to the causality constraints inherent in the temporal character
of the contact network: for instance, if node $i$ interacts first with node $j$
and then with node $k$, a message or infectious agent
can travel from $j$ to $k$ through $i$, but not in the opposite direction,
while in a static network both events would be equally possible.

It is therefore interesting to study the spreading paths of the SI
process on a temporal network and on the corresponding aggregated network,
as mentioned in the above Section~\ref{sec:si}. 
To this aim, for each seed node we define a \textit{transmission network}
along which the infection effectively spreads in the temporal network
(i.e., the network whose edges are given by $S\leftrightarrow I$ contacts).
Therefore, the distance along the transmission network
between the seed node and another arbitrary node $i$
gives the actual number of transmission events that occurred
before the spreading process reached $i$,
and consequently it is the length of the \textit{fastest} path
from the seed individual to the infected one
which respects the causality constraints of the temporal network \cite{Moody:2002,kleinberg:2008,Bajardi:2011}.

\begin{figure}[ht]
\centering
\includegraphics[width=\columnwidth]{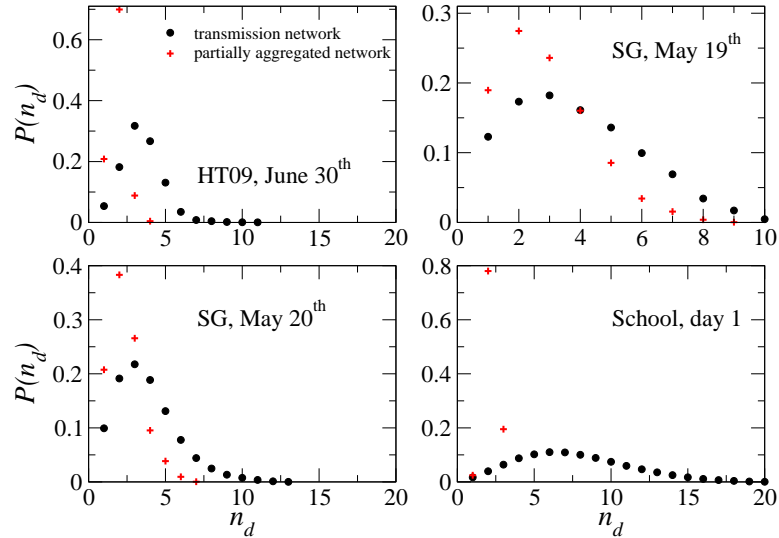}
\caption{Distribution of the path lengths $n_{d}$
from the seed node to all the infected individuals,
computed over the transmission network (black circles)
and over the partially aggregated networks (red pluses).
For each day the distributions are computed
by varying the choice of the seed node over all individuals.}
\label{path-length-distr}
\end{figure} 

On the other hand, in an aggregated, static view of the contact network
the spreading would follow the \textit{shortest} path over
the network aggregated from the time the seed first appears,
as the infection can only spread along interactions occurring after the arrival of the seed node.
We call this network a \textit{partially aggregated network}.

As shown in Figure~\ref{path-length-distr},
the distribution of lengths of the fastest paths turns out to be broader and shifted
toward higher values than the corresponding shortest-path distributions.
This holds both for the conference and for the museum case,
and has also been observed in other cases~\cite{kleinberg:2008}.
The actual number of intermediaries is therefore larger on a temporal network
than would be predicted by a propagation scheme based on a static network.
This difference can be understood by considering a similar example as discussed above:
if node $i$ is infected and interacts with node $j$, who then interacts with node $k$
before $i$ interacts with $k$, the actual (fastest) spreading path between nodes $i$ and $k$
has path length $2$, while the shortest path has a unitary path length.

In settings in which each transmission event has a cost,
or is associated with the possibility of signal loss or attenuation,
such differences might play an important role and
temporal effects should accordingly be taken into account carefully.

\subsection{Activity clocks}
\label{sec:intrinsic}
\begin{figure}
\includegraphics[width=.45\columnwidth]{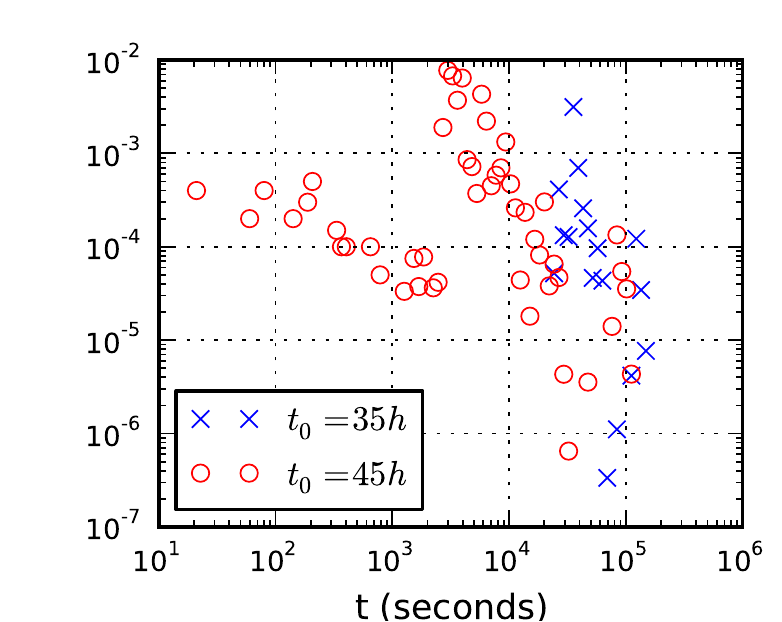}
\includegraphics[width=.45\columnwidth]{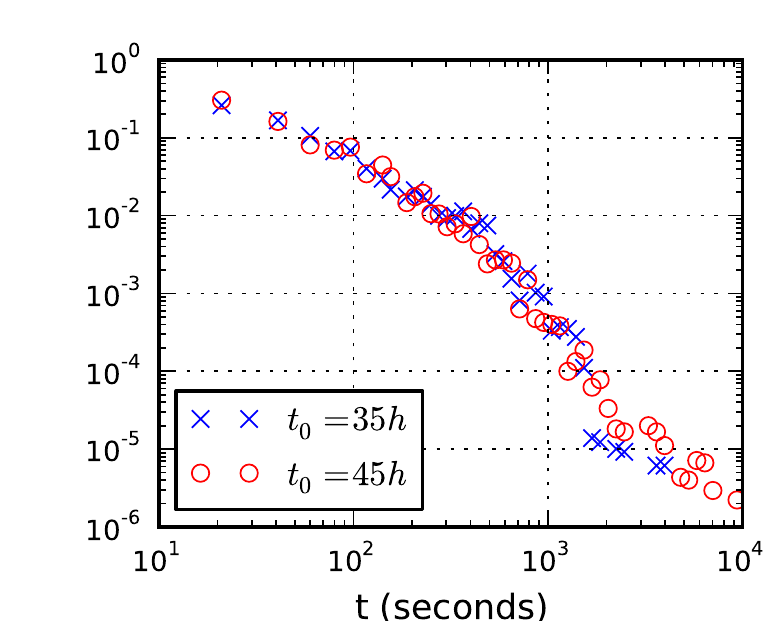}
\caption{Log-binned distribution of message arrival delays.
The distributions are computed by simulating a simple SI process
over the HT09 temporal contact network, for two different injection times
of the initial message (blue crosses and red circles).
Left: the delay interval is defined as the wall-clock time difference between
the arrival of the message at a given node and the injection time of the message.
Right: the delay interval distribution is computed by using node-specific
activity clocks that only run when nodes have active links (proximity relations)
to other nodes. That is, each node has a specific notion of ``intrinsic'' time
that is defined as the cumulated time it spent in contact with other nodes.
}
\label{figpanisson}
\end{figure}

The results of the previous subsection show that,
depending on the environment and on the time at which the spreading process is initiated,
different spreading dynamics can be obtained.
In particular, the time at which a given node is reached by the information or infection
may strongly depend on collective activity patterns.
In the context of message routing in networks of mobile devices~\cite{Hui:2005, Zhang:2007, Boldrini:2008, Lee:2009},
the distribution of elapsed times between the generation of the message and its arrival at given nodes
is often considered as a way to evaluate the performance of a spreading protocol~\cite{Groenvelt:2005, Cai:2007, Miklas:2007, Karvo:2008}.
However, the non-stationary and bursty behavior of the contact and proximity networks
imply that the distribution of delays between message injection and message delivery at a given node
may in fact depend importantly on the time of injection of a message,
or on specific details of contact and activity patterns.
The left panel of Fig.~\ref{figpanisson} shows this for the case of a message
that spreads over the HT09 temporal contact network according to a simple SI process
initiated at two different points in time:
the distribution of arrival delays computed in terms of wall-clock time
is extremely sensitive to the injection time of the message
and displays strong heterogeneities that cannot be captured
by any simple statistical model. 
It is thus important to devise more robust metrics for message delivery
that factor out non-stationary behaviors and temporal heterogeneities, allowing
on the one hand to carry out more objective comparisons of different protocols for message spreading,
and on the other hand to validate the models of human mobility (and the ensuing temporal networks)
that are used to design and inform such protocols.

Some progress in the direction outlined above can be made
by giving up the global notion of wall-clock time in favor of a node-specific definition of time.
We imagine that each node has its own clock,
and that this clock only runs when the node is involved in one or more contacts.
Since the clock measures the amount of time a given node has spent
in interaction, we refer to this clock as an ``activity clock''.
All activity clocks are set to zero at the beginning of the spreading process,
when the initial message is injected into the network.
Thus, the activity clock of a node measures the amount of time during which that node
could have received a message propagated along the links of the temporal network,
i.e., it ignores the time intervals during which the node was disconnected from
the rest of the network.

In this context, the message delivery delay for node $i$ is defined as the value
of its activity clock when the message is received,
i.e., it is the elapsed time node $i$ has spent in contact with others,
from the injection time of the initial message
up to the moment when the message is received by $i$.
The right panel of Fig.~\ref{figpanisson}
shows the distributions of delivery delays computed in terms of activity clocks.
The distributions exhibit a smooth dependence on activity-clock time,
without the strong heterogeneities observed when using wall-clock time.
Most importantly, they now collapse onto one another,
i.e., they are robust with respect to changes in the injection time of the message.
Strikingly,  they are also robust with respect to the context:
as reported in Ref.~\cite{Panisson:2011},
the same distribution is obtained for spreading processes
simulated in conferences with very different schedules and contact densities.

It is important to remark that synthetic temporal networks of human contact
are typically generated on the basis of a number of accepted models
for the underlying human mobility in space. The quality of the models
is often assessed on the basis of their ability to reproduce simple
statistical observables of the temporal network, such as the probability distribution
of contact durations and the distribution of times between successive contacts of a node.
However, when computing the distributions of delivery delays for an SI process
in terms of activity clocks, strong differences emerge between the accepted
generative models and the empirical temporal networks. Specifically,
as reported in Ref.~\cite{Panisson:2011}, most of the accepted models
yield delay distributions that are much narrower than those of Fig.~\ref{figpanisson},
thus failing to reproduce the empirical phenomenology.
This points to the need for further modeling work to design generative models of temporal
networks that can correctly reproduce the empirical distributions.
Notice that such differences only become visible on using activity clocks
in combination with dynamical process on the temporal network, used as a probe
of properties that cannot be captured by means of customary metrics.
For example, the comparison between the left and right panels of Fig.~\ref{figpanisson}
shows how suitable definitions of ``time'' based on activity metrics
of the temporal network have the power to uncover regularities
that are otherwise completely shadowed by the intrinsic heterogeneities of the system.
At the same time these results call for more research in several directions.
In particular, it would be important to:
\begin{itemize}
\item define and characterize activity clocks based on different types of node and edge metrics for temporal networks (e.g., defining time as the cumulated number of contact events, rather than as the elapsed time in contact),
\item express in terms of activity clocks the evolution of simple dynamical processes taking place on temporal networks, 
\item model the form of the delay distributions observed in simulation for paradigmatic dynamical processes such as SI spreading taking place on temporal networks.
\end{itemize}

\section{Conclusion}
\label{sec:conclusion}

The study of empirical temporal networks is receiving increasing attention
because of the availability of new data sources and because of their relevance
for the detection, modeling and control of phenomena in a broad variety of
socio-technical systems. To date, key questions are still open
about temporal networks and dynamical processes over temporal networks.

Here we focused on high-resolution empirical temporal networks of human proximity,
provided a phenomenological overview of some important properties,
and reported on the impact that such properties have on paradigmatic
epidemic-like processes that take place over temporal networks,
with potential applications to diverse domains such as information spreading or epidemic modeling.
We remark that the consolidated toolbox of statistical indicators, network metrics
and generative models for static networks that has been developed over the last decade
cannot be trivially generalized to the case of temporal networks. 
More research is thus needed in order to identify dynamical extensions of the observables
designed for static network, to uncover dynamical network motifs \cite{Bajardi:2011,kovanen},
and to explore entirely new characterization techniques
that expose important features of the temporal-topological structure of the networks.
In this respect, it will be probably fruitful to investigate simple
dynamical processes over networks as a tool for uncovering relevant patterns,
and as an aid to constrain and evaluate generative models for temporal networks.

We also notice that in the context of many applicative domains it is very difficult to characterize
separately the dynamics of the network and that of the relevant dynamical process
(e.g., the dynamics of a contact network and the dynamics of an epidemic unfolding over it):
depending on the specific process, on its timescales, and on the set of properties
of the process that one aims at modeling, the temporal structure of the network
can have a profoundly different relevance. 
For example, it has been shown that for flu-like epidemic processes,
with latent and infectious timescales of the order of a few days,
the temporal structure of the underlying human contact networks
is negligible if one aims at modeling just the peak time of the epidemics,
and is important if the goal is to model the size of the epidemics~\cite{Stehle:2011b}.
Developing a more systematic understanding of the impact
that the temporal structure of a network has on a given observable is an open challenge.

The above remarks are related to another important problem, that of coarse-graining
temporal network data when either explicit node/edge attributes are available
or node/edge activity patterns can be clustered into classes of similar behavior.
When dealing with human contact networks, this is the case of many contexts
in which the population under study is structured because of roles (e.g., hospitals)
and/or spatio-temporal constraints on group mobility and interactions (e.g., schools).
In particular, inferring behavioral classes from temporal network data
requires algorithms to mine for dynamical communities of nodes or edges
that extend those available for static networks, as a temporal network
may have sharply defined classes of dynamical behavior that completely
disappear on considering aggregated networks obtained by projecting out time.
To this end, machine learning techniques based on node/edge features
or on entire activity timelines may prove effective in uncovering
and characterizing behavioral regularities hidden in temporal networks.
When classes are known (e.g., explicit role-based classes within a hospital population),
or discovered via time-aware community detection techniques,
it is often insightful to consider aggregated representations of the data
based on the class attributes, such as the contact matrices commonly
used in epidemiology. These customary representations, though, have been
defined and investigated in order to coarse-grain static interaction networks,
and they need to be generalized so that they can be used to summarize
temporal networks in a way which is suitable for the relevant applicative context.

Progress in the above directions of creating synopses of temporal network data
would greatly help in defining ``parsimonious'' models for dynamical processes,
that only retain the necessary amount of information about the underlying temporal network.
This also calls for work on reconciling different scales of representation
of temporal network data, so that properties and dynamical processes
at different levels can be related to one another. 
In consideration of the coming deluge of behavioral information
represented in the form of temporal networks, the ability to create parsimonious but informative
representations will be an increasingly valuable asset for the applications of network science.

\acknowledgement{ It is a pleasure to thank G. Bianconi, V. Colizza,
  L. Isella, A. Machens, A. Panisson, J.-F. Pinton, M. Quaggiotto,
  J. Stehl\'e, W. Van den Broeck, A. Vespignani for many interesting
  discussions. We also warmly thank all the collaborators who helped
  make the SocioPatterns deployments possible, and in particular
  Bitmanufaktur and the OpenBeacon project. Finally, we are grateful
  to all the volunteers who participated in the deployments.  }


\bibliographystyle{epj}

\end{document}